\documentclass{aa}
\usepackage{psfig}

\makeatletter
\def\@citex[#1]#2{\if@filesw\immediate\write\@auxout{\string\citation{#2}}\fi
\def\@citea{}\@cite{\@for\@citeb:=#2\do
{\@citea\def\@citea{,\penalty\@m\ }\@ifundefined
{b@\@citeb}{\@warning
{Citation `\@citeb' on page \thepage \space undefined}}%
{\csname b@\@citeb\endcsname}}}{#1}}
\makeatother
\let\internalcite\cite
\def\cite{\def\citename##1{##1}\internalcite}
\def\ctyr{\def\citename##1{}\internalcite}

\def\bron{\object{1E 2259+58.6}}
\def\Xray{\mbox{X-ray}}

\begin{document}
\thesaurus{06(08.16.7 \bron; 02.01.2; 08.14.1; 08.23.1; 13.25.5)}

\title{A deep search for the optical counterpart to the anomalous X-ray
pulsar \bron} 
%
\author{F. Hulleman\inst{1}
   \and M. H. van Kerkwijk\inst{1}
   \and F. W. M. Verbunt\inst{1}
   \and S. R. Kulkarni\inst{2}}
\institute{Astronomical Institute, Utrecht University, 
           P. O. Box 80000, NL-3508 TA~~Utrecht, The Netherlands
   \and    Palomar Observatory, California Institute of
           Technology 105-24, Pasadena, CA 91125, USA}
\offprints{F. Hulleman}
\mail{F.Hulleman@astro.uu.nl}

\date{Received January 10 2000, accepted }
\maketitle

\begin{abstract}
We present Keck R and I band images of the field of the anomalous
\Xray\ pulsar \bron.  We derive an improved X-ray position from
archival ROSAT HRI observations by correcting for systematic
(boresight) errors.  Within the corresponding error circle, no object
is found on the Keck images, down to limiting magnitudes $R=25.7$ and
$I=24.3$.  We discuss the constraints imposed by these limits, and
conclude that it is unlikely that \bron\ is powered by accretion from
a disk, irrespective of whether it is in a binary or not, unless the
binary is extremely compact.

\end{abstract}

\section{Introduction}
\label{intro}

There is a growing number of \Xray\ pulsars for which it is not clear
why they emit \Xray{}s.  Their rotational energy loss appears
insufficient to power the \Xray\ luminosity and there is no sign of a
companion from which they could accrete matter.  These pulsars are
called the ``anomalous X-ray pulsars'' (AXP).

The AXP differ from other X-ray pulsars in the following properties
(\cite{Hellier}, \cite{MS95}, \cite{Paradijs}): (i) their spin periods
are within a small range, between 5 and 9 seconds, and their spin-down
rates are nearly constant; (ii) timing analysis gives no evidence of
orbital motion; (iii) their \Xray\ spectra are rather soft; (iv) their
\Xray\ light curves show less variability; (v) their
inferred \Xray\ luminosities are rather low
($\sim\!10^{35}\,\mathrm{erg\,s^{-1}}$); (vi) they are located close to
the Galactic plane; (vii) some appear associated with supernova
remnants.

Members of the group of AXP include \object{4U 0142+61}, \bron,
\object{1E 1048.1$-$5937} and \object{RX J1838.4$-$0301} (\cite{MS95}
and references therein).  Recently, three \Xray\ pulsars with
similar properties were discovered: \object{1RXS J170849.0$-$400910}
(\cite{Sugizaki}); \object{1E 1841$-$045} in the supernova remnant
\object{Kes 73} (\cite{Kes73}); and \object{AX J1845.0$-$0300}
(\cite{Torii}; \cite{GV}) in the supernova remnant G29.6+0.1
(\cite{GGV99}). Indeed, perhaps even the progeny of these
systems has been identified, in \object{RX J0720.4$-$3125}
(\cite{Haberl}).  The latter object is the only one for which the
optical counterpart has been identified (\cite{MH98}; \cite{KvK98}).
 
A variety of models for AXPs has been put forward, using both
accretion and intrinsic mechanisms to explain the \Xray\ luminosity.
For the accretion scenarios, the source of matter has been suggested
to be a very low mass companion (\cite{MS95}) or a circumstellar disk
composed of debris, either from a common envelope phase of a high mass
\Xray\ binary (\cite{Paradijs}; \cite{Ghosh}) or from a supernova
explosion (\cite{Corbet}; \cite{Chatterjee}).  In scenarios in which
the \Xray\ emission is intrinsic, the sources are either hot, massive,
rapidly spinning white dwarfs, formed in the merger of a double
white-dwarf binary (\cite{Paczynski}), or so-called ``magnetars,''
isolated neutron stars with very high magnetic field strengths
($\ga\!10^{14}\,\mathrm{G}$; \cite{TD96}).  In the latter case, AXP
might be related to soft gamma-ray repeaters\footnote{The proposed
\Xray\ counterparts to SGRs also have spin periods in the 5--9 s range
(\cite{SGR1806}, \ctyr{SGR1900}).}.

In this paper, we present a deep search for the optical counterpart to
the AXP \bron.  In Sect.~\ref{sec:bron}, we briefly review what is
known about this source.  We derive a refined position for \bron\ in
Sect.~\ref{sec:position}, and in Sect.~\ref{sec:keck} we present
our optical observations.  We discuss the results in
Sect.~\ref{sec:discuss}.

\section{\bron}
\label{sec:bron}

\bron\ is located inside the supernova remnant \object{CTB 109}
(\object{G109.1$-$1.0}).  This remnant is semi-circular, with neither
\Xray\ nor radio emission from its western side (see,
e.g., \cite{RP97}), which may be related to the presence of a
molecular cloud.  A lobe of enhanced \Xray\ emission (`jetlike'
feature) is present in between the eastern shell and the pulsar, but
likely this does not reflect physical interaction.

\bron\ is an \Xray\ pulsar with a period of 6.98\,s.  It is observed
to be spinning down with a period derivative of
$5.5\times10^{-13}\,\mathrm{s\,s^{-1}}$.  Baykal \& Swank
(\ctyr{Baykal}) found small but significant deviations from constant
spin-down and concluded that accretion must be the source of energy
(but see Melatos \ctyr{Melatos} for a different explanation).  An
upper limit $a_\mathrm{X}\sin i<0.03$ lt-s has recently been obtained
with the RXTE (\cite{MSI98}).  Assuming a Roche-lobe filling companion
and an orbital inclination $i>30\degr$, this restricts the companion
mass to $\la\!0.1M_{\sun}$ for a main-sequence star, and to
$\la\!0.8M_{\sun}$ for a helium-burning star.  No constraint can be
set for a white-dwarf companion.

\bron\ has a very soft \Xray\ spectrum, best described by a
combination of a power-law with photon index $4.0\pm0.1$ and a
blackbody with $kT=0.43\pm0.02\,\mathrm{keV}$, absorbed by a column
density $N_\mathrm{H}=8.5\times10^{21}\,\mathrm{cm^{-2}}$ (\cite{RP97}
using ASCA, BBXRT and ROSAT PSPC data; see also \cite{Corbet}).  A
possible problem is contamination of the spectrum by extended emission
around the pulsar (\cite{RP97}).  Parmar et al.\ (\ctyr{Parmar}) tried
to account for this by subtracting the observed spectrum of the
supernova remnant multiplied with a scale factor.  They find that the
residual spectrum can be fitted with a simple power law with an
absorption column density of
$N_\mathrm{H}=(2.18\pm0.07)\times10^{22}\,\mathrm{cm^{-2}}$.  It
should be noted, however, that the BeppoSAX LECS and MECS instruments
used by these authors have rather poor spatial resolution.
Furthermore, Rho \& Petre (\ctyr{RP97}) found that the spectral shape
of the extended emission near \bron\ differs from that of the remnant.

No radio counterpart to \bron\ has been found down to an upper limit
of $50\,\mathrm{\mu{}Jy}$ (\cite{CJL94}).  So far, no counterpart at 
optical and infrared wavelengths could be identified (\cite{DC91};
\cite{CP98}). These authors obtained limiting magnitudes of B $=25$,
V $=24$, R $=24.5$, I $=23$, H $=18.5$, J $=19.6$ and K $=18.4$, although
they could not exclude several candidates due to the fact that the
\Xray\ position of \bron\ was not known accurately enough. Note 
that their astrometry differs from ours and that of Fahlman
et~al. (\ctyr{Fahlman82}) (see Sect.~\ref{sec:position} and Davies \& Coe
\ctyr{DC91}).    

\subsection{Distance and reddening to \bron}
\label{sec:red}

All distance estimates to \bron\ use distances to either \object{CTB
109} or the molecular cloud, i.e., it is assumed that the pulsar is at
the same distance as \object{CTB 109} and that the molecular cloud is
either at the same distance as well or in front (and causes the absence of
emission from the western part of the \object{CTB 109}).  The distance
to \object{CTB 109} was estimated using the surface brightness
diameter relation by Sofue et~al. (\ctyr{Sofue}) and Hughes
et~al. (\ctyr{Hughes}), at 4.1 and 5.6\,kpc, respectively.  A lower
limit to the distance of $d\ga6\pm1\,\mathrm{kpc}$ was obtained by Strom
(priv.comm) from hydrogen 21-cm line absorption measurements towards the
remnant (\cite{Braun}).

For the molecular cloud, a kinematic distance of $5\pm1\,\mathrm{kpc}$
can be derived from the molecular line measurements of Kahane
et~al. (\ctyr{Kahane}) (here, we used the linear fit to the galactic
rotation curve of Fich et~al. \ctyr{FBS89}, with rotation
constants $\Theta_0=220\,\mathrm{km\,s^{-1}}$ and
$R_0=8.5\,\mathrm{kpc}$).  Distance estimates have also been made for
several \ion{H}{ii} regions associated with the molecular cloud.
Spectrophotometric parallaxes for the exciting stars were obtained by
Crampton et al.\ (\ctyr{Crampton}), leading to distances of 5.5, 5.6,
3.6 and 4.0\,kpc for the \ion{H}{ii} regions S148, S149, S152 and
S153, respectively.  Furthermore, a lower limit of
$d\ga6\pm1\,\mathrm{kpc}$ towards S152 was obtained by Strom (priv.comm)
using hydrogen 21-cm line absorption.  From the above, assuming that
the fractional errors in the spectrophotometric parallaxes are about 25\%
(corresponding to uncertainties of about 0.5\,mag in the absolute
magnitudes inferred from the spectral types), we conclude that a
distance to the molecular cloud of $5\pm1\,\mathrm{kpc}$ is most
likely, and, therefore, that \bron\ is at $d\ga5\pm1\,\mathrm{kpc}$.


The reddening to \bron\ can be estimated from the \Xray\ hydrogen
column density using the relation
$N_\mathrm{H}=(1.79\pm0.03)\times10^{21}\,A_V\,\mathrm{cm^{-2}}$,
where $A_V$ is the visual extinction in magnitudes (\cite{PS95}).  From
the fits to the \Xray\ spectrum of Rho \& Petre (\ctyr{RP97}), values
of $N_\mathrm{H}$ between 0.8 and $1.2\times10^{22}\,\mathrm{cm^{-2}}$
are found, with a best fit value of
$0.85\times10^{22}\,\mathrm{cm^{-2}}$.  Thus, one infers $A_V=4.5$ to
$6.7\,$mag, with a best-fit value of 4.7\,mag.  For
$N_\mathrm{H}=2.18\times10^{22}\,\mathrm{cm^{-2}}$, as found by Parmar
et~al.\ (\ctyr{Parmar}), one would infer $A_V=12.2\,$mag.  As noted
above, however, we feel this measurement is rather uncertain.
 
The reddening can be estimated indirectly using \object{CTB 109}.
Fesen \& Hurford (\ctyr{Fesen}) measured H$\alpha$ and H$\beta$
relative line intensities in optical spectra of 5 filaments, and found
$A_V=2.5$ to $3.7\,$mag, with substantial variations across the
remnant.  These values appear somewhat lower than those derived above
from the \Xray\ absorption for \bron, and thus one may wonder about the
association, or about possible intrinsic absorption around \bron.
Since the reddening to the remnant is seen to vary, however, a
somewhat higher value for \bron\ does not seem unlikely.
Furthermore, from the \Xray\ emission of \object{CTB 109}, column
densities corresponding more closely to that of \bron\ are inferred
(\cite{RP97}).

In the following, we will adopt a reddening of $A_V=4.7\,$mag and a
distance of 6 kpc, unless noted otherwise.  

\section{Improved X-ray position}
\label{sec:position}

\bron\ has been in the field of view of two observations with the
ROSAT high resolution imager (HRI; David et~al. 1999). The first was
in 1992 from January 8, 07:16~UT, to January 10, 18:59~UT, and was
centered on \bron.  The second was pointed at the lobe to the East of
the pulsar and was from June 25, 07:46~UT, to June 27,
08:26~UT.  In both observations, the semicircular shell, the lobe, the
pulsar and the extended emission in its immediate surroundings, as
well as three other point sources can be seen.  We identify the latter
with sources 1, 2 and 3 in Table~4 of Rho \& Petre (\ctyr{RP97}), and
will refer to these as RP1, RP2 and RP3 hereafter.

The data were analysed using the EXSAS software package (1998 April
version) in the following way.  First, we corrected for the small
error in HRI plate scale found by Hasinger et~al. (\ctyr{Has98}), by
multiplying all detector positions (relative to the detector centre)
with a factor 0.9972.
Next, we ran the standard source detection program, which creates a
preliminary source list using a sliding cell algorithm.  This source
list is fed into a maximum likelyhood detection algorithm
(\cite{Cruddace}), which yields precise positions (see
\cite{Zimmermann}).  In order to prevent trouble with the extended
emission from the remnant, we used two masks, one to the East of
\object{CTB 109} containing RP3 and one to the West containing RP1,
RP2 and \bron.  Our final positions are listed in table~\ref{tab02};
the 1-$\sigma$ errors are internal, i.e., they do not yet include
pointing uncertainties.  We verified that different choices of mask
images gave consistent results.

Comparing the positions derived for the two observations
(Table~\ref{tab02}), it is clear that they are not consistent.  This
reflects the uncertainties in the attitude solutions used to transform
detector positions to sky positions (see, e.g., \cite{David}).  One
can correct for this if one has independent positions for some of the
sources, for instance from optical counterparts.  Below, we will try
to do this.

\begin{table}[t]
\caption[]{\Xray\ and optical positions used for the astrometry.}
\begin{tabular}{llll}
\hline
Source & $\alpha_\mathrm{J2000}$ &
$\delta_\mathrm{J2000}$&$\sigma_{\alpha,\delta}{}^\mathrm{a}$\\ 
\hline
\multicolumn{2}{@{}l}{\em\object{RP1} -- HRI}\\
January 1992&               23 00 32.85&  58 52 51.9&  0.9\\
June 1992&                  23 00 33.98&  58 52 49.5&  2.3\\
\multicolumn{2}{@{}l}{\em\object{RP1} -- counterpart}\\
1425-14436845$^\mathrm{b}$& 23 00 33.396& 58 52 47.16& 0.2\\
DSS1 1953.830&              23 00 33.393& 58 52 47.22& 0.2\\
DSS2 1991.671&              23 00 33.377& 58 52 47.07& 0.2\\[2mm]
\multicolumn{2}{@{}l}{\em\object{RP3} -- HRI}\\	 
January 1992&               23 03 19.41&  58 45 33.9&  1.3\\
June 1992&                  23 03 20.30&  58 45 26.4&  1.5\\
\multicolumn{2}{@{}l}{\em\object{RP3} -- counterpart}\\
1425-14515707$^\mathrm{b}$& 23 03 19.472& 58 45 28.95& 0.2\\
DSS1 1953.830&              23 03 19.471& 58 45 28.43& 0.2\\
DSS2 1991.671&              23 03 19.578& 58 45 28.68& 0.2\\[2mm]
\multicolumn{2}{@{}l}{\em\bron\ -- HRI}\\
January 1992&               23 01 08.09&  58 52 48.3&  0.1\\
June 1992&                  23 01 08.99&  58 52 45.3&  0.1\\
\multicolumn{2}{@{}l}{\em\bron -- corrected}\\
Using RP1 and RP3&                23 01 08.44&  58 52 44.1&  0.9\\
Using RP1 only&                   23 01 08.62&  58 52 43.5&  1.3\\
\hline
\end{tabular}
\label{tab02}
\\$^\mathrm{a}$ Uncertainty in each coordinate (in arcsec).
\\$^\mathrm{b}$ USNO-A2 identifier; mean epoch 1953.830.
\end{table}

Many X-ray sources can be identified with bright stars; for instance,
Motch et al.\ (\ctyr{Motch}) find that 85\% of the sources in a
galactic plane region are bright stars.  Therefore, we searched the
USNO-A2.0 catalogue (\cite{usnoa2}) for bright stars
($r_\mathrm{usnoa2}<13.0$) near the positions of RP1, RP2 and RP3.  We
tentatively identify RP1 and RP3 with stars \object{USNO-A2
1425-14436845} and \object{USNO-A2 1425-14515707}, respectively.  No
bright star was found within $30\arcsec$ of the position of RP2.  We
derived proper motions for the two stars by measuring positions on
first and second generation images from the digitized sky survey,
using astrometry relative to the USNO-A2.0 catalogue\footnote{Note
that both DSS1 and USNO-A2.0 positions were derived from the same plate
and the results should therefore agree. The slight discrepancy in
declination for RP3 is probably due to a second (fainter) star close
to the proposed counterpart. This does however not affect our results.}
(Table~\ref{tab02}).  For \object{USNO-A2 1425-14515707}, we found a
significant proper motion of
$0\farcs022\pm0\farcs004\mathrm{\,yr^{-1}}$ at position angle
$90\fdg7\pm10\fdg2$. 

The refined position for \bron\ was derived as follows.  First, the HRI
positions were shifted to a common reference using the positions of
\bron.  Next, we determined weighted average offsets in right ascension and
declination between the HRI and optical positions (including proper
motion).  For this purpose, we weigh the four individual offsets in
right ascension and declination (RP1 and RP3; two HRI observations)
using the uncertainties given by the source detection algorithm (the
errors in the optical positions are much smaller, $\sim\!0\farcs2$).

The differences with the weighted average correspond to
$\chi^{2}=11.5$ for 6 degrees of freedom.  This is not a good fit and
suggests the presence of additional systematic errors.  Following
Hasinger et~al.\ (\ctyr{Has98}), we added an error $E_\mathrm{sys}$ in
quadrature to the individual measurement errors such that
$\chi^{2}_\mathrm{red}\simeq1$.  For this purpose, we require
$E_{sys}=1\farcs0$.  Including this additional uncertainty, we infer
that our corrected position of \bron, listed in Table~\ref{tab02},
has an uncertainty of 0\farcs9 in each coordinate.  This corresponds
to a 95\% confidence error radius of 2\farcs2.

The value of $E_\mathrm{sys}$ we find is twice the value used by
Hasinger et al.\ (\ctyr{Has98}).  While these authors also included an
$0\farcs5$ uncertainty in optical positions, we wondered whether the
somewhat large required value might be related to the rather large
offset from the center of RP3.  To get a feeling for the uncertainty,
we also determined a corrected position using RP1 only.  The position
is shifted slightly and has a slightly larger uncertainty of 1\farcs3
in each coordinate (see Table~\ref{tab02}).  The 95\% confidence error
circle for this case has a radius of~3\farcs3.

We show both error circles superposed on Keck images in
Fig.~\ref{figkeck}.  Also indicated are the positions determined
previously by Fahlman et~al. (\ctyr{Fahlman82}; short dashed) and
Hanson et~al. (\ctyr{Hanson}; long dashed).  Note that all these
positions differ significantly from that of Coe \& Pighting
(\ctyr{CP98}).  We do not understand the reason for this discrepancy.

\section{Keck Observations}
\label{sec:keck}

\begin{figure*}[t]
\psfig{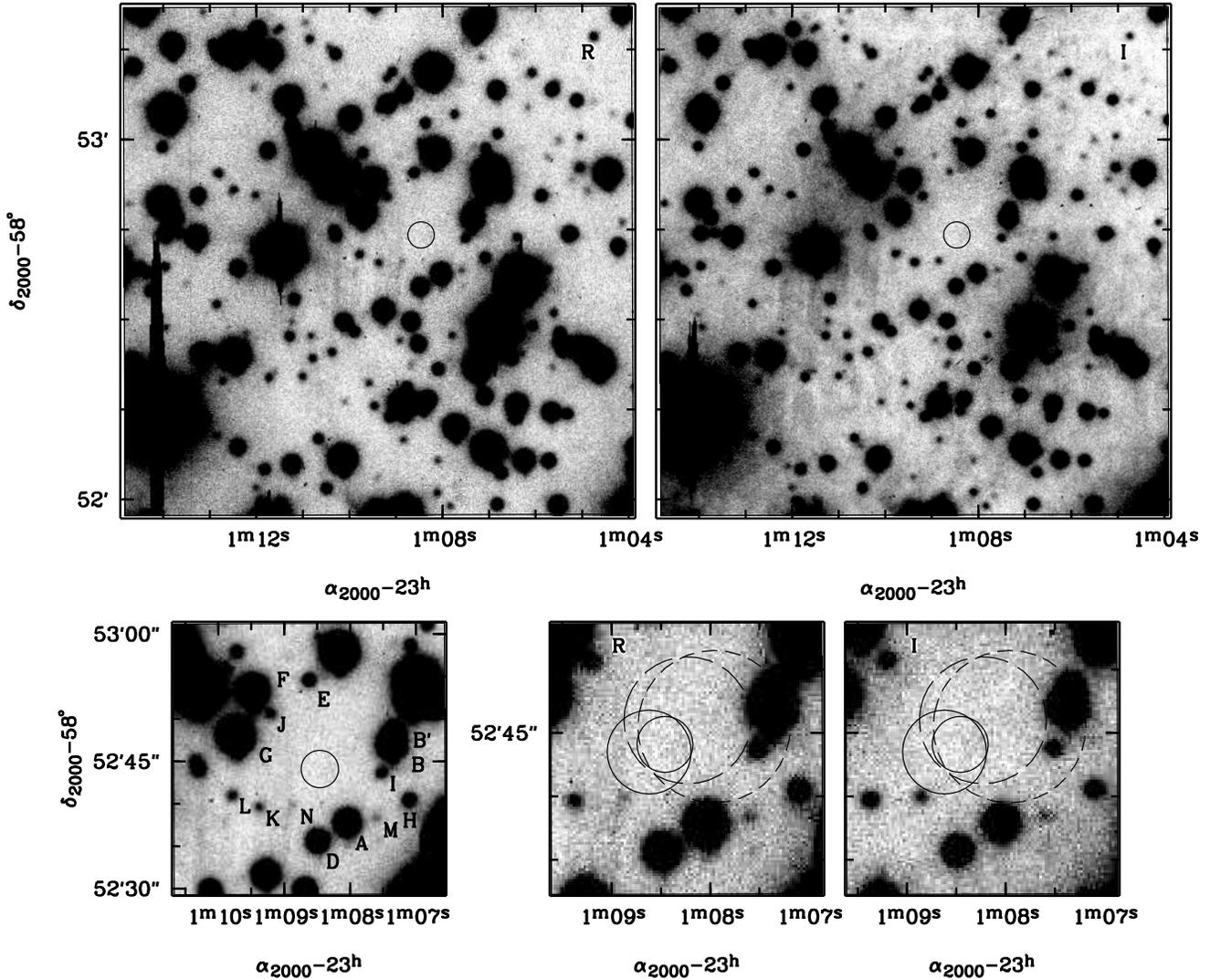}
\caption[]{R and I band images around \bron, the circle is the 95\%
confidence circle obtained in Sect.~\ref{sec:position}. This circle
is also shown in the lower three
enlarged images, where one has the star identifications and two have the
conservative Rosat error circle at 95\% confidence and previous error
circles obtained by Fahlman et~al. (\ctyr{Fahlman82}; short dashed)
and Hanson et~al. (\ctyr{Hanson}; long dashed) superposed. 
\label{figkeck}}
\end{figure*}

Optical (R and I band) images of the field around \bron\ were obtained
on 1994 October 30/31 and 1997 January 7 using the Low-Resolution
Imaging Spectrometer (LRIS, \cite{Oke}) at the Cassegrain focus of the
Keck~I (1994) and Keck~II (1997) telescopes.  The 1994 observations
were plagued by clouds and high cirrus; the seeing was 1\farcs1.  All
1994 images have a large gradient in the sky background.  As a result,
all but one of these images are useless.  In 1997, the weather was
photometric, but the seeing mediocre at 1\farcs2.  All images were
bias subtracted and flat fielded using dome flats.  Unfortunately, it
turned out that in the 1997 R-band images the position of \bron\ was
very close to a region of lower sensitivity in the CCD, which did not
flat-field out.  The resulting ``smear'' is just over stars K and N in
two of the three images; it can also be seen in
Fig.~\ref{figkeck}).

\begin{table}[t]
\caption[]{Apparent magnitudes$^\mathrm{a}$ of selected stars in the
field around \bron. Star designations are after Davies \& Coe (\ctyr{DC91}).}
\begin{tabular}{cccr}
\hline
Star & $R$  & $I$ & $R-I$\\
\hline
A & $19.697\pm0.003$ & $18.577\pm0.005$ & $1.120\pm0.006$\\
B & $19.265\pm0.003$ & $18.217\pm0.005$ & $1.048\pm0.006$\\
B' & $21.206\pm0.010$ & $19.987\pm0.011$ & $1.219\pm0.015$\\
D & $20.409\pm0.004$ & $19.274\pm0.005$ & $ 1.135\pm0.006$\\
E & $22.537\pm0.016$ & $22.545\pm0.043$ & $-0.008\pm0.046$\\
F & $18.902\pm0.004$ & $17.904\pm0.007$ & $0.998\pm0.008$\\
H & $22.398\pm0.014$ & $21.219\pm0.014$ & $1.179\pm0.020$\\
I & $23.024\pm0.023$ & $21.672\pm0.020$ & $1.352\pm0.030$\\
J & $23.131\pm0.027$ & $21.473\pm0.019$ & $1.658\pm0.033$\\
K & $23.676\pm0.037$ & $21.297\pm0.015$ & $2.379\pm0.040$\\
L & $23.389\pm0.028$ & $21.799\pm0.028$ & $1.590\pm0.040$\\
M & $24.557\pm0.086$ & $22.832\pm0.068$ & $1.745\pm0.110$\\
N & $24.848\pm0.110$ & $22.119\pm0.031$ & $2.729\pm0.114$\\
\hline
\label{tab01}
\end{tabular}
\\\noindent $^\mathrm{a}$ In addition to the listed uncertainties,
there is a 0.1\,mag uncertainty in the zero point; see text.
\end{table}

Astrometry was done using a short (300\,s) R band image from 1997.  We
identified objects with stars in the USNO-A2.0 catalogue and obtain
centroids for objects that were not saturated and appeared stellar
(i.e., had a sufficiently small diameter).  Next, we applied a
bi-cubic distortion correction (J.~Cohen, 1997, private comm.), and
solved for zero-point position, scale, and position angle by
comparison with the USNO-A2.0 positions.  After rejecting six outliers
(probably misidentifications or large proper motion objects), the
remaining 134 objects gave a good solution. The root-mean-square displacement
from the catalogued positions is 0.26\arcsec\ in right ascension and
0.25\arcsec\ in declination.   

For the photometry, we used the DAOPHOT II software package
(\cite{Stetson}).  Instrumental magnitudes were obtained from two R
band images (one from 1994 and one from 1997, 1000s exposure each) and
the average of two I band images (300s each, 1997 only).  We
calibrated the photometry using the magnitudes determined by Davies \&
Coe (\ctyr{DC91}).  The apparent magnitudes of selected objects as
determined from the 1997 observations are listed in Table~\ref{tab01}.
The values derived for the 1994 R-band image are consistent with
these, but have lower accuracy.  Note that the uncertainties listed in
the table are the formal $1\sigma$ fitting errors.  In addition to
these, there are systematic uncertainties related to the zero point.
Since Davies \& Coe (\ctyr{DC91}) found a discrepancy by about 0.15
mag in magnitudes obtained from two different exposures, the
zero-point uncertainty is probably of the order of 0.1\,mag.

No object is found inside either of the HRI error circles derived
above.  We derived lower limits with the help of simulations.  For
these, we took the mean of the three 1997 R band images and selected
the inner region of the image that is void of stars.  Next, we placed
an artificial star with magnitude $M$ at a random spot in this region
and determined its magnitude in the same way as for the real objects
(this also allowed us to verify the errors quoted in the table).  This
was done ten times for a number of values of $M$, within a range
25.0--26.9 for R and 23.7--25.7\,mag for~I.  We define the $3\sigma$
detection threshold to be that magnitude $M$ for which the standard
deviation equals 0.3 mag.  Taking account of the fact that we need
about 3 trials to cover the error circle, the derived $2\sigma$ limits
are $R>25.7$ and $I>24.3$ (for the error circle derived using RP1 only
we need 7.5 trials and obtain $R>25.6$ and $I>24.2$).

\section{Discussion}
\label{sec:discuss}  

We now examine the limits on the optical emission
imposed by the various models proposed for \bron. Contributions to the
optical flux can come from the object itself, an accretion disk and a
companion star. We will discuss these, in the context of the proposed
models, in the following subsections.

\subsection{An isolated object}
\label{sec:merged}

Paczy\'{n}ski (\ctyr{Paczynski}) proposed that \bron\ is a rapidly
rotating, hot, massive, hot white dwarf, formed in the merger of two
normal white dwarfs.  Such a rapidly rotating white dwarf can have
sufficient rotational energy loss
($\sim10^{36}\mathrm{\,erg\,s^{-1}}$) to power the observed \Xray\
flux.  Using blackbody emission and a white dwarf radius of
$0.002\,R_{\sun}$, our optical limits correspond to a maximum
temperature $kT=0.4\,$keV.  Since the Eddington condition gives a
maximum temperature of $kT=0.2\,$keV (\cite{Paczynski}), our data
can thus not exclude this model.

Similarly, our limits do not allow us to constrain the magnetar
model.  (A very crude estimate gives $R\simeq37$ for a $kT=0.4\,$keV
blackbody at $d=6\,$kpc and reddened by $A_R=3.8$.)

\subsection{An isolated NS accreting from a disk}

\begin{figure}[t]
\psfig{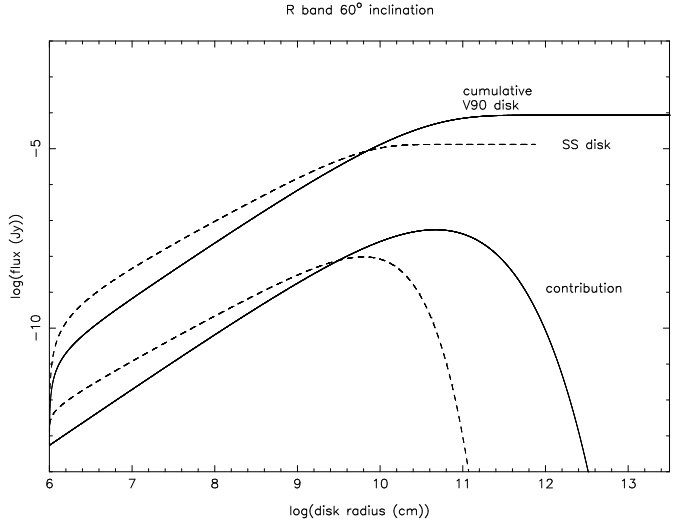}
\caption[]{Cumulative R band flux observed at earth as a function of
radius for irradiated (solid line) and unirradiated (dashed line)
disks at a distance of 6 kpc.  The lower curves show the contribution
of the flux for individual disk annuli (at constant $\Delta\log{}r$).
No correction was made for reddening.
\label{fig:disk}}
\end{figure}

We will make an estimate for the optical emission of an accretion disk
using two simple models.  In the first, we use the
$T(r)\propto{}r^{-3/7}$ relation expected for an accretion disk
dominated by irradiation (labeled V90 hereafter; \cite{Vrtilek}).  An
important assumption in obtaining this relation is that the disk is
isothermal in the vertical direction, due to the irradiation.  The
second model we use is a Shakura-Sunyaev disk (labeled SS;
\cite{SS73}), in which the energy generated by the disk itself is
taken into account but irradiation is neglected.  For both models, we
assume the disk emits like a blackbody.  We believe the first model
should be the more reasonable approximation (at least for radii larger
than a few $10^{10}\,$cm; for smaller radii, this model underestimates
the emission; see Fig.~\ref{fig:disk}), while the second model should
give a conservative limit.

Fig.~\ref{fig:disk} shows predicted R-band fluxes as a function of
disk radius for a disk inclined by 60\degr\ relative to the line of
sight, located at a distance of 6\,kpc, and either irradiated by a
central \Xray\ source with
$L_\mathrm{X}=4.3\times10^{35}\,\mathrm{erg\,s^{-1}}$ (V90), or not
irradiated (SS).  Also shown is the contribution to the total flux
from each radius $r$ (at constant $\Delta\log{}r$).  There is no
significant contribution to the flux for radii larger than
$\sim\!10^{12}\,$cm (V90) and $\sim\!10^{11}\,$cm (SS), because the
temperature becomes too low.  For a large disk around an isolated
neutron star, we find $R=22.7$ for the V90 disk ($r\geq10^{12}\,$cm;
$i=60\degr$), and $R=24.7$ for the SS disk ($r\geq10^{11}\,$cm;
$i=60\degr$).  Here, for both estimates we used an R-band extinction
$A_R=3.8$, which corresponds to $A_V=4.7$ (see Sect.~\ref{sec:red}).

From the above limits, we conclude that it is unlikely that \bron\ is
an isolated NS accreting from a disk.  Only if we underestimated the
reddening considerably or if the disk inclination were very
unfavourable, a (large) accretion disk could have escaped detection.
Increasing the distance to the source is not an option, since the
optical emission of the disk is proportional to the \Xray\ emission.
(for the SS disk model at 10\,kpc, we still expect $R=25.1$). 

\subsection{A very low mass binary}

The optical emission of low mass \Xray\ binaries is dominated by the
emission from the accretion disk surrounding the compact object. An
empirical relationship,
$M_\mathrm{V}=(1.57\pm0.24)-(2.27\pm0.32)\times
\log[(L_\mathrm{X}/L_\mathrm{edd})^{1/2}(P/1\,\mathrm{hr})^{2/3}]$,
was found by van~Paradijs \& McClintock (\ctyr{vPenMcC}).  Assuming
$(V-R)_0\sim0$, according to this relationship a putative binary must
have $P\la10^{2}\,$s in order to remain below our detection limit
$R>25.7$.  This would rule out all but a white dwarf counterpart.
There is considerable scatter around the relation, however, and period
estimates will be off by a factor of 5 for a change of 1\,mag.
Furthermore, it is far from clear one can extrapolate their result to
such short periods.


Given the above, it is perhaps more relevant to compare our limits
with the emission from the compact low-mass \Xray\ binary \object{4U
1626$-$67} (\cite{Chakrabarty}).  Indeed, this source has a number of
the properties of AXP (\cite{MS95}; cf.~\cite{Paradijs}).  Under the
assumption that \object{4U 1626$-$67} and \bron\ have the same
$L_\mathrm{X}/L_\mathrm{opt}$ ratio, the expected magnitude would be
$R=24.8$.  Here, we have taken account of the difference in \Xray\
flux of about a factor ten (compare, e.g., \cite{Orlandini} and
\cite{Parmar}), as well as the difference in reddening towards the two
sources.  Even if the \Xray\ flux of \bron\ were only one twentieth
that of \object{4U 1626$-$67}, we would still expect to have detected
the optical counterpart at $R=25.5$.  Thus, we conclude that most
likely even such a compact binary would have been detected.

What would our limits imply if there were no accretion disk present
and the optical emission was dominated by the companion star?  This
seems unlikely, but might perhaps be the case if accretion is from a
(irradiation-induced) stellar wind.  We will discuss two possible
companions below: (i) a main-sequence star; (ii) a He-burning star.
From Sect.~\ref{sec:merged}) above, it will be clear that we can put
no meaningful constraints for the case the only source of optical
emission is a white-dwarf companion.

Based on the colors derived from model spectra by Bessell
et~al. (\ctyr{Bessell}) (overshoot models, their Table~1, for
$\log{}g=4.5$), we find that a main-sequence companion to \bron\ would
have to be of type K or later to escape detection.

To start helium burning, a star's mass must exceed
$\sim\!0.3M_{\sun}$.  From Savonije et~al.\ (\ctyr{Savonije}), we find
that a $\sim\!0.3M_{\sun}$ helium burning star has
$T_\mathrm{eff}\simeq20000\,$K and $L\simeq0.6L_{\sun}$.  Using
Table~3 of Bessell et~al.\ (\ctyr{Bessell}), we infer a bolometric
correction of $-2.6$ and $V-R=-0.1$ for a 20000\,K star (note that
this table does not extend to $\log{}g>5$, but since the variation
with $\log{}g$ is small, our extrapolation should be safe).  For these
values, the star would have $R=25.7$ and $I=24.7$, which is just below
our detection limit.  More massive stars would be brighter, and, thus,
if the companion is a star burning helium it must have a mass close to
the limit of $0.3\,M_{\sun}$.
 


\section{Conclusions}

We have shown that the tight limits we derive on the optical emission
of \bron\ are a serious problem for scenarios involving accretion by
way of an accretion disk.  This is irrespective of whether the source
is in a binary, except if the binary is extremely compact. Therefore,
we conclude that, most likely, \bron\ is an isolated object, and that
its emission is not powered by accretion. We note that the same
conclusion was indepently drawn by Kaspi et~al. (\ctyr{Kaspi}) on the
basis of the extreme rotational stability of \bron.

\begin{acknowledgements}
We would like to thank Richard Strom for discussions about the
distance to \bron.  We have made use of the ROSAT Data Archive of the
Max-Planck-Institut f\"{u}r extraterrestrische Physik (MPE) at
Garching, Germany, of the SIMBAD database operated at CDS, Strasbourg,
France, and of the Digitized Sky Surveys, which were produced at the
Space Telescope Science Institute from photographic data obtained
using the Oschin Schmidt Telescope on Palomar Mountain.  We
acknowledge support of a fellowship of the Royal Netherlands Academy
of Arts and Sciences (MHvK) and grants from NASA and NSF (SRK).
\end{acknowledgements}

\end{document}